\documentclass[showpacs,showkeys,11pt,
preprint,preprintnumbers,nofootinbib,
groupedaddress,superscriptaddress,amsmath,amssymb]{revtex4}
\usepackage{amsfonts}
\usepackage{amsmath}
\usepackage{graphics}
\usepackage{epsfig}
\usepackage{url}
\usepackage{multirow}
\usepackage{feynmp}
\usepackage{xcolor}

\newcommand {\be}{\begin{equation}}
\newcommand {\ee}{\end{equation}}
\newcommand {\ba}{\begin{eqnarray}}
\newcommand {\ea}{\end{eqnarray}}

\begin{document}
\title{Charged Higgs observability in the single top production channel at LHC}

\author{Nadia Kausar}
\email{nkausar430@gmail.com}
\author{Ijaz Ahmed}
\email{Ijaz.ahmed@cern.ch}
\affiliation{Riphah International University, Sector I-14, Hajj Complex, Islamabad Pakistan}


\begin{abstract}
 Single top quark production through weak interactions is  an important source of charged Higgs in the Minimal Super Symmetric Standard Model.  In the s-channel single top production ,charged Higgs having largest cross-section may appear as a propagator in the form of heavy resonance state decaying to a pair of top and bottom quark. In this paper the channel under consideration is $pp \rightarrow H^{\pm}\rightarrow tb \rightarrow b\bar{b}W^{\pm} \rightarrow b\bar{b}\tau^{\pm} \nu_{\tau}$, where top quark exclusively decays into a pair of b quark and W boson while W boson subsequently decays to $\tau$ jet and neutrino. So the final state is characterized by the presence of two b jets, hadronic $\tau$ decay and missing transverse energy.It has been demonstrated that the charged Higgs signal observability is possible within the available MSSM parameter space (tan$\beta$, $m_{H^{\pm}})$ respecting all experimental and theoretical constraints due to presence of QCD multijet and electroweak background events at LHC. In order to show the observability potential of charged Higgs, the charged Higgs signal are well observed or excluded in a wide range of phase space particular ($\beta > 35$ at 500 $fb^{-1}$  , $\beta > 25$ at 1000 $fb^{-1}$ , $\beta > 15$ at 3000 $fb^{-1}$ ) at $\sqrt{s}=$14 TeV.

\end{abstract}
\maketitle
\section{Introduction}
It has been seen that the discovery of the neutral scalar particle at Large Hadron Collider (LHC) with a measured mass of 125 GeV approximately by the CMS and ATLASexperiments\cite{Higgs_Discovery1,Higgs_Discovery2,Higgs_Discovery3} proved that the Standard Model (SM) predictions are well compatible with the measured values. However, there are very convincing evidences from experimental signatures and theoretical calculations that SM needs to be overlooked by some other dynamical models in order to deal with the issues regarding neutrino masses, dark matter and the hierarchy problem. Several attempts were made to extend the SM scalar sector which  resulted in the form of Minimal Supersymmetric Standard Model (MSSM) \cite{mssm1,mssm2,mssm3},Two Higgs Doublet Model (THDM) \cite{2hdm1,2hdm2,2hdm3,2hdm4}  and several others\cite{nmssm1,nmssm2}.Among these physics models several are testable with extended Higgs sector at LHC In this study the MSSM framework is used as a benchmark which leads to five physical scalar Higgs bosons: light Higgs boson h,  heavy Higgs boson with CP-even H, a CP-odd Higgs boson A and two charged Higgs bosons $H^{\pm}$. The discovery  of these new Higgs particles would be a clear indication that the extended Higgs sector to be a source of Electroweak Symmetry Breaking. \\
Apart from other Higgs particles, the charged Higgs has significantly interesting and challenging features at particle colliders, because it provides unique signature due to having charge which provides different kinds of signatures in terms of their interactions and decay properties from other neutral Higgs bosons. If $m_{H^{\pm}} < m_{t} - m_{b}$ , the dominant mechanism for charged Higgs production is via top quark decay: $t \rightarrow bH^{\pm}$. Therefore the charged Higgs are produced preferably via $t\bar{t}$ production. The focus of this study at Tevatron, LEP and LHC was mostly performed in the domain of low mass charged Higgs, where charged Higgs predominantly decays into a pair of $\tau$ lepton and neutrino at tan$\beta > 5$ \cite{feynhiggs1,feynhiggs2,feynhiggs3} or into pair of charm and strange jets $(H^{\pm}\rightarrow cs)$. If $m_{H^{\pm}} > m_{t} + m_{b}$ ,then the dominant production  mode is  top quark associated production $H^{\pm}tb$, where charged Higgs prefer to decay into a pair of top and bottom quark, i.e., $H^{\pm}\rightarrow tb$. This decay $H^{\pm}\rightarrow tb$ is  kinematically allowed,  an interesting channel for heavy charged Higgs analysis. Identification of the signals $t\bar{t}b\bar{b}$ becomes hard due to presence of huge irreducible background signals .   That is why the early analyses usually focused on the decays like $H^{\pm}\rightarrow \tau \nu$ and $H^{\pm}\rightarrow cs$  in order to use  the advantage of suppressed background signals by using $\tau$-identification techniques. \\
In the past few years, the charged Higgs study was also proposed at LHC through single top production processes in both light and heavy mass regions and it  resulted in quite promising agreements for discovery of  charged Higgs . Currently, there are just a few analyses where single top acts as a source of charged Higgs boson. For light charged Higgs study the decay of top quark is $(pp\rightarrow tq\rightarrow qbH^{\pm}\rightarrow qb\tau\nu)$ in the t-channel single top production ,if  $\tau$ lepton decays hadronically \cite{st1} or leptonicaly \cite{SingleTop_CH}. For heavy charged Higgs study the leptonic final state is $(pp\rightarrow tb\rightarrow bbW^{\pm}\rightarrow bbl^{\pm}\nu_{l})$ in s-channel single top production \cite{st2}.  The contribution of off-diagonal coupling between the two incoming quarks like cb in the production cross-section of s-channel charged Higgs may lead to enhance the total cross-section  by a factor of 2.7 \cite{myplb} . Similarly  \cite{st3}  the single top production is considered  a source of charged Higgs exchange in t-channel , through being observable at very high integrated luminosities andhigher values of  tan$\beta$ . Recently, the fully hadronic final state for charged Higgs is studied  to investigate the potential of the charged Higgs discovery in all jets final state in the s-channel  \cite{ijazepjc}. This paper will follow the continuation of the same process butwith $\tau$-jet final state the $\tau$-jet identification algorithm is applied in an alternate way for reconstruction of charged Higgs   . The objective of this article is the reconstruction of charged Higgs invariant mass in s$-$channel single top production in chain $(pp\rightarrow tb\rightarrow bbW^{\pm}\rightarrow b\bar{b}\tau^{\pm}\nu_{\tau})$, where two b-jets and one $\tau$-jet along with missing transverse energy are founded at LHC. This work is accomplished using Monte Carlo generators performing hadronization and fragmentation processes and concentrating on the charged Higgs mass in the available region of the parameter space not yet excluded, +i.e., $200 <m_{H^{\pm}}<400$ GeV. The exclusion bounds and discovery reach will be explored in the context of  2HDM Type II which is a special case of MSSM. There are some background processes like W+jets and QCD multi jets that could make it a challenging analysis. However, by applying some kinematic cuts wisely, they can be well  controlled.\\
The current experimental constraints on charged Higgs searches have been performed at colliders like LEP  \cite{lepexclusion1}, Tevatron \cite{cdf3} and LHC using the ATLAS \cite{CHtaunu8TeVATLAS,CHIndirect8TeVATLAS,LightCH7TeVATLAS2,LightCH7TeVATLAS1} and CMS \cite{cms2,cms3} experiments. In the direct and indirect searches, the experimental exclusion limits are set by these experiments and can be seen in \cite{ijazepjc}.\\
The following sections of the article contain signals and background processes generation which naturally shows  interference and event selection procedure where different mass hypotheses are presented for reconstruction of charged Higgs invariant mass. A 5 $\sigma$ discovery and exclusion at 95$\%$ Confidence Level (CL) is provided in the accessible regions of MSSM parameter space. The $m_h-max$ scenario with the given parameters: $M_{2}$ = 200 GeV,~$M_{\tilde{g}}$ = 800 GeV,~$\mu$ = 200 GeV and $M_{SUSY}$ = 1 TeV is used and estimation of all parameters is performed within theoretical framework of MSSM.The given scenario $m_h-max$  defines a benchmark point optimized to  maximize the upper bound on $m_{h}$ for a fixed $m_{t}$, the soft SUSY breaking parameter $M_{SUSY}$ at a given tan$\beta$.This benchmark point gives the maximum value of parameter space in the direction of  $m_{h}$ and conservative exclusion limits for tan$\beta$. \\

\subsection{Collider Analysis}

The signal process under study is $pp \rightarrow tb\rightarrow W^{\pm}bb\rightarrow \tau\nu bb$, where the top and bottom quarks are produced through the s-channel and charged Higgs are  mediating as resonance in the single top production process. The $\tau$ lepton is analyzed through its hadronic decay. The main background processes with same final states arises from processes i.e., $W^{\pm}jj$, $W^{\pm}bb$, $W^{\pm}cc$, single top t-channel and $t\bar{t}$. We study the fully hadronic decay of $\tau$ lepton which   esults from the W boson decay. The decay widths and cross-sections are extracted as calculated in \cite{ijazepjc}. The parton density function is provided by LHAPDF 5.9.1 \cite{lhapdf} with the version CTEQ 6.6. The background processes $W^{\pm}jj$, $W^{\pm}bb$, $W^{\pm}cc$ combinely referred as "W + 2jets" are generated with Madgraph5 2.3.3 \cite{madgraph}     version with a kinematic preselection cut applied on jets as $E_{T}^{jets}>20$ GeV and $|\eta|<3.0$, while single top t-channel, $t\bar{t}$ samples are produced by PYTHIA 8.1.5.3 \cite{pythia}. All the signal processes are produced with CompHEP 4.5.2 \cite{comphep}. The output of both these packages in Les Houches Event Format (LHEF) is used by PYTHIA8 for partonic showering, gluon radiation, fragmentation and hadronization. The reconstruction of all jets is performed with the jets clustering algorithm FASTJET 3.1.3 \cite{fastjet} using anti-kt algorithm \cite{antikt} and $E_{T}$ recombination scheme. The jet cone size is fixed at $\Delta R=0.4$, where $\Delta R=\sqrt {(\Delta \eta)^{2}+(\Delta \phi)^{2}}$ is jet cone radius, $\phi$ is azimuthal angle and $\eta =-ln$tan$\theta/2$ is pseudorapidity.
 
\subsection{Event Selection strategy}
\begin{figure}[tb]
 \centering
   \includegraphics[width=0.6\textwidth]{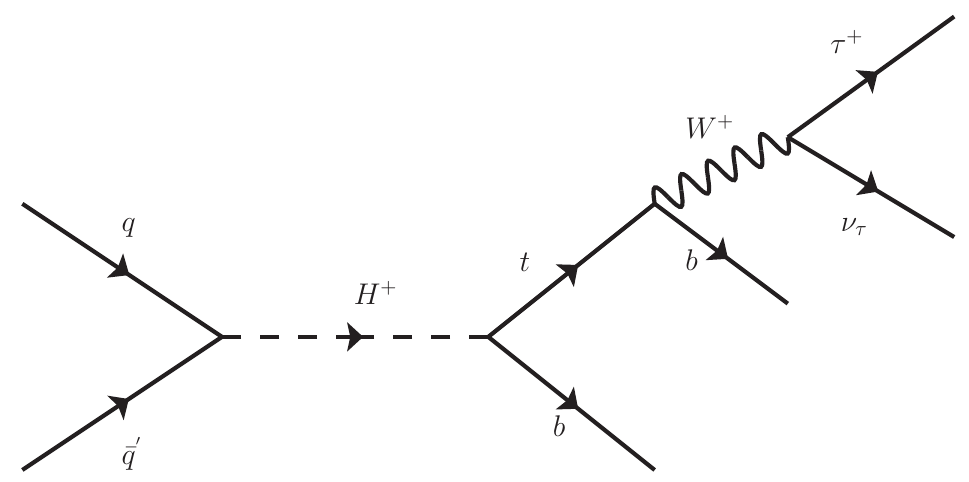}
   \caption{The s-channel single top production via charged Higgs as a signal resonance process with $\tau$ hadronic decays.}
   \label{diagram}
 \end{figure}

\begin{figure}[tb]
 \centering
   \includegraphics[width=0.6\textwidth]{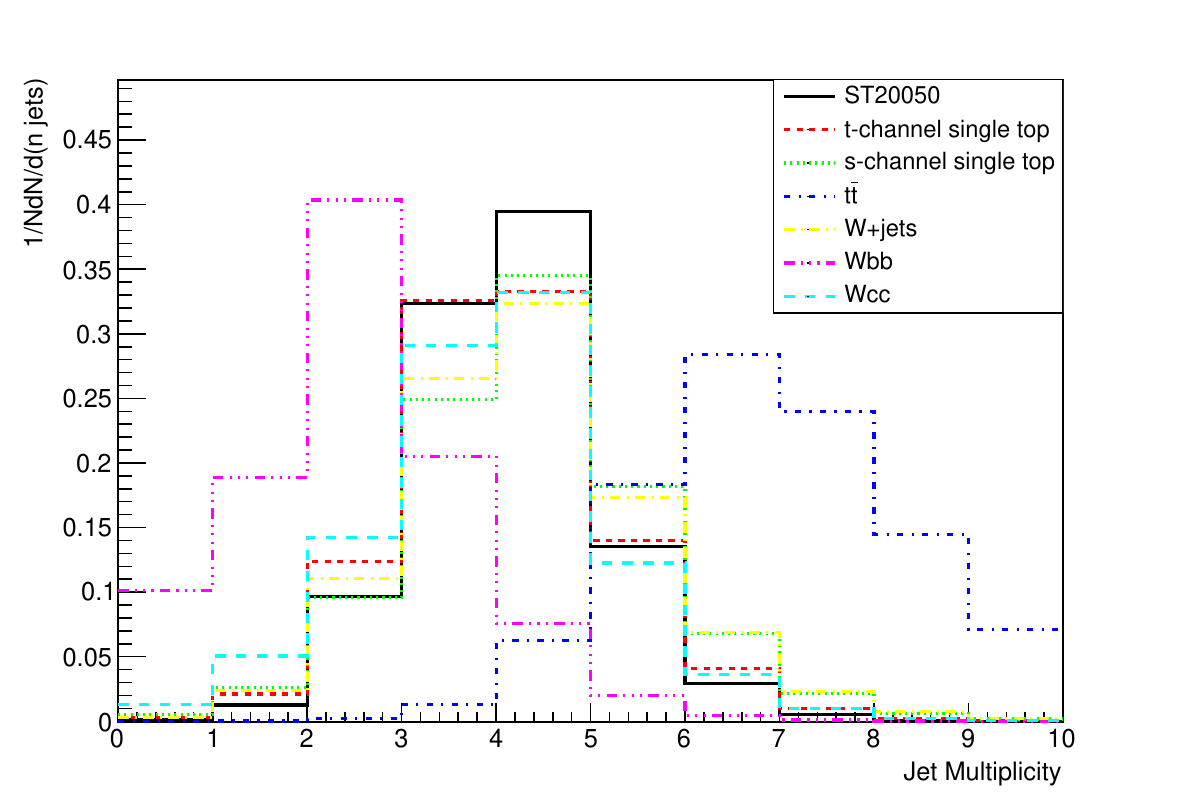}
   \caption{The jet multiplicity is shown with both signal and background events for a center of mass energy of 14 TeV. .}
   \label{alljets}
 \end{figure}

A reasonable and effective  understanding of event kinematics is required for  the suppression of background events and signal events. The topological and kinematical differences between signal and background events help to apply the selection cuts. By plotting the kinematic distributions the above goals are achieved.  The final state topology of the signal process under consideration contains two b-jets and One $\tau$-jet along with missing transverse energy $E_{T}$ as shown in Figure~\ref{diagram}. All the selection cuts are shown in Table~\ref{selectioncuts}.
The existance of two b-jets in the events is expected to suppress the large "W + 2jets" samples. However $\bar{t}t$ will be seen to be main background at the end. The signal consists of a charged Higgs with mass $m_{H}= 200$ GeV and tan$\beta=$ 50 areis shown through out the paper. The tan$\beta$ only contributes to the signal cross-section and also it is irrelevant for such distributions that is why the abbreviation used for the signal is “ST20050".Here 200 is mass of charged Higgs and 50 is value of $\tan\beta$.  This selected $\tan\beta$ and $m_{H^{\pm}}$ value has the highest cross-section. In Figure~\ref{alljets} jets multiplicity is shown with  missing transverse energy $E_{T}>20$ GeV and $|\eta|<3$. If the total number of jets are equal to three \textcolor{red}{,} the events are selected to perform further processes.
 The reconstruction of b-jets is performed by using the jet-parton matching algorithm   where a jet is tagged to be a b-jet if the $\Delta R$ value between the jet and bottom or charm quarks with $p_{T}>20$ GeV and $|\eta|<3$ is less than 0.4. The efficiency of b-jet identification is assumed to be 40$\%$ while the c-jet mis-tagging rate is assumed to be 10$\%$.
\begin{figure}[tb]
 \centering
   \includegraphics[width=0.6\textwidth]{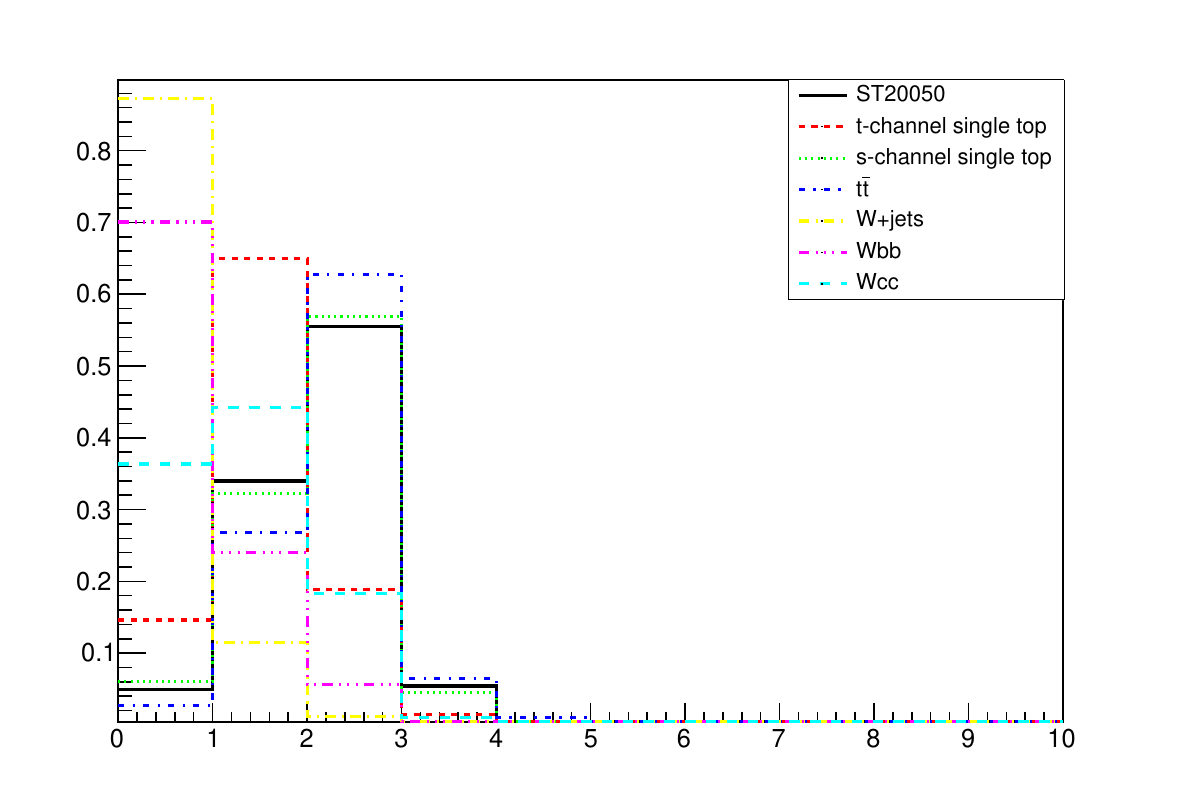}
   \caption{The multiplicity distribution of b-jets with both signal and background events are shown.}
   \label{bjets}
 \end{figure}

\begin{figure}[tb]
 \centering
   \includegraphics[width=0.6\textwidth]{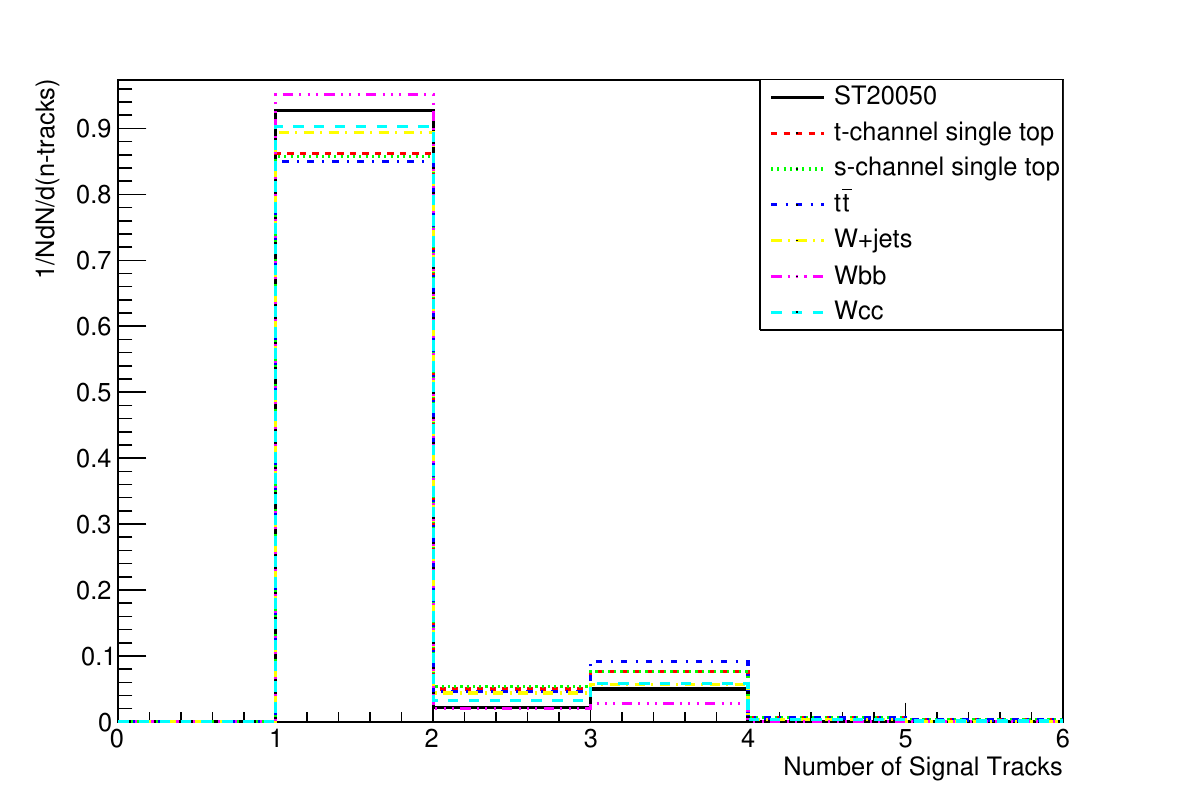}
   \caption{Number of tracks in the signal cone of $\tau-$jet candidate in signal as well as background events}
   \label{nstracks}
 \end{figure}

\begin{figure}[tb]
 \centering
   \includegraphics[width=0.6\textwidth]{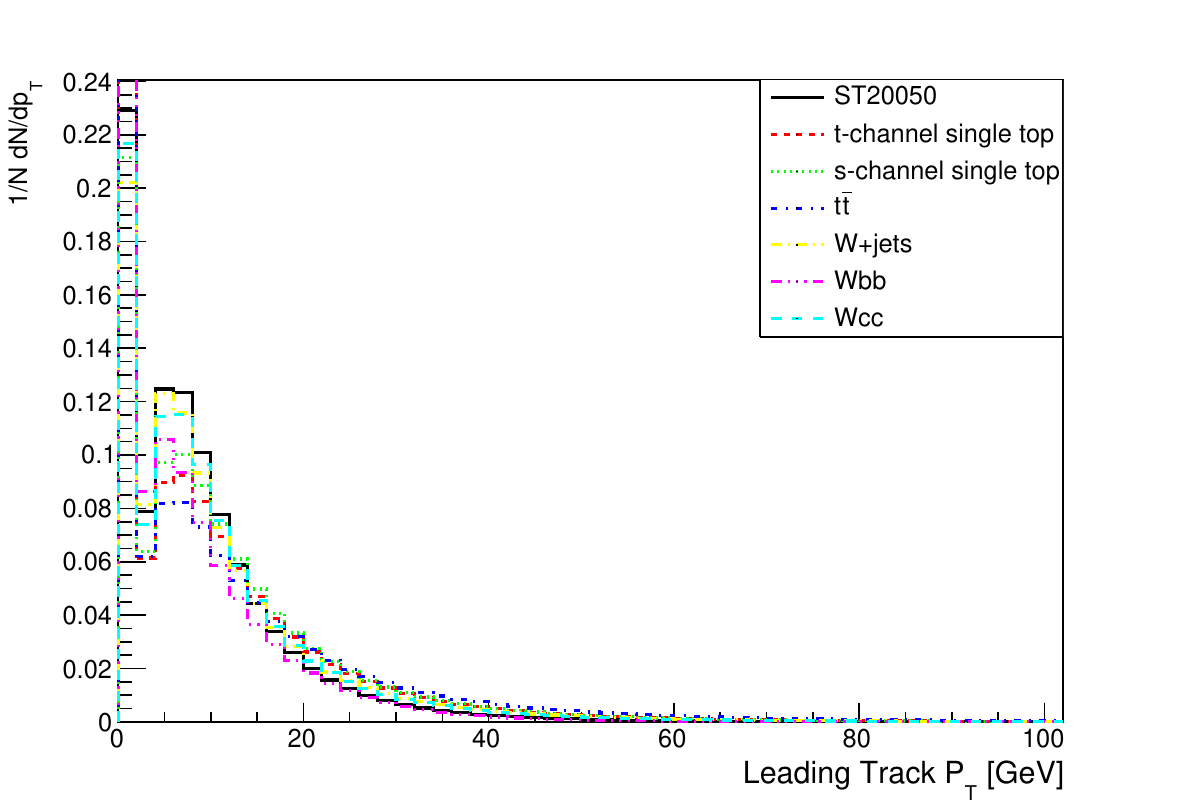}
   \caption{Leading tracks distribution in both signal and background events.}
   \label{LeadingTPt}
 \end{figure}

\begin{figure}[tb]
 \centering
   \includegraphics[width=0.6\textwidth]{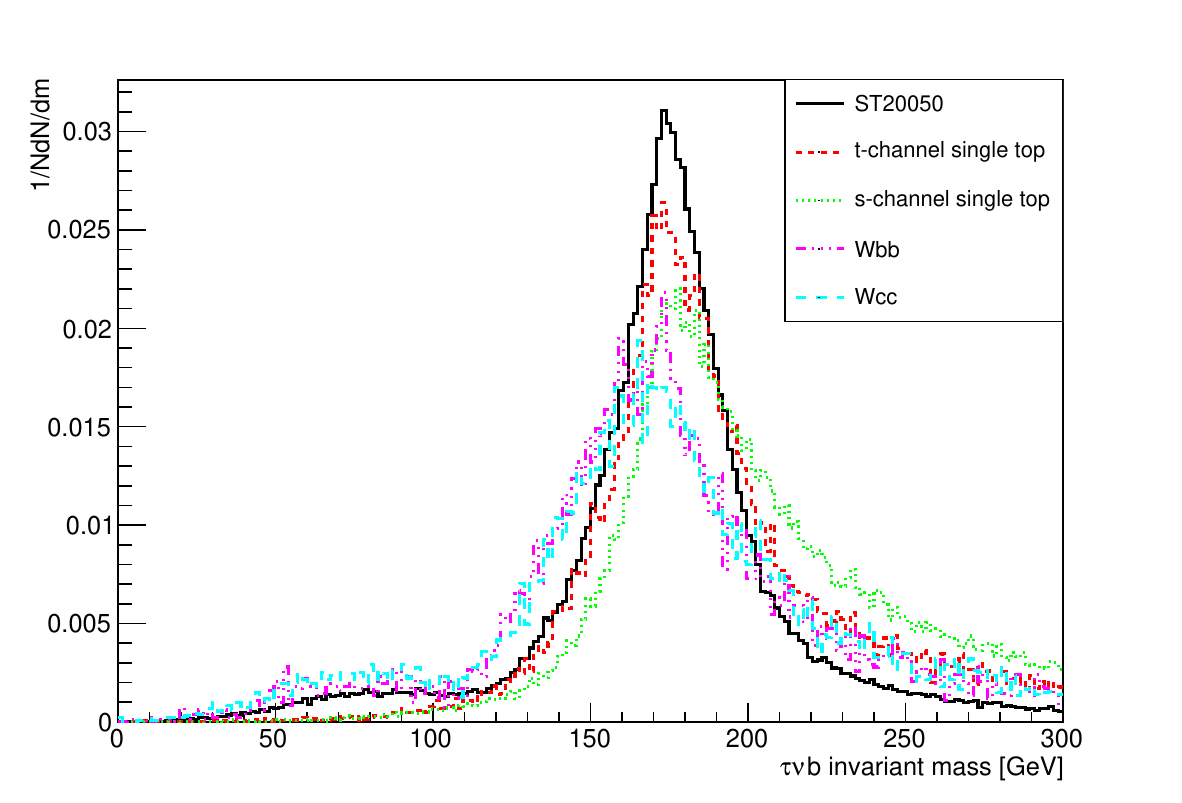}
   \caption{The reconstructed invariant mass distribution of $b\tau\nu$ candidate as signal and dominant background.}
   \label{topmass}
 \end{figure}

\begin{figure}[tb]
 \centering
   \includegraphics[width=0.6\textwidth]{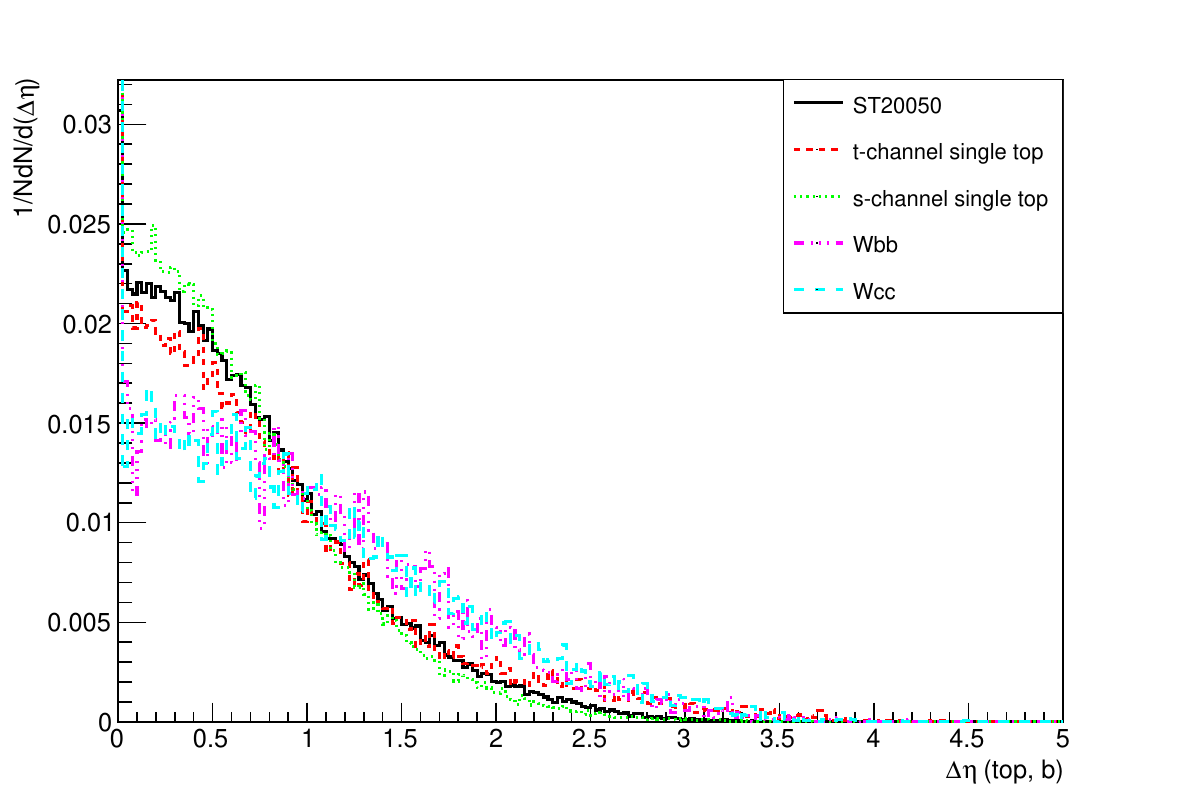}
   \caption{The pseudorapidity difference between the tau-jet and bottom quark is shown. }
   \label{deltaeta}
 \end{figure}

\begin{figure}[tb]
 \centering
   \includegraphics[width=0.6\textwidth]{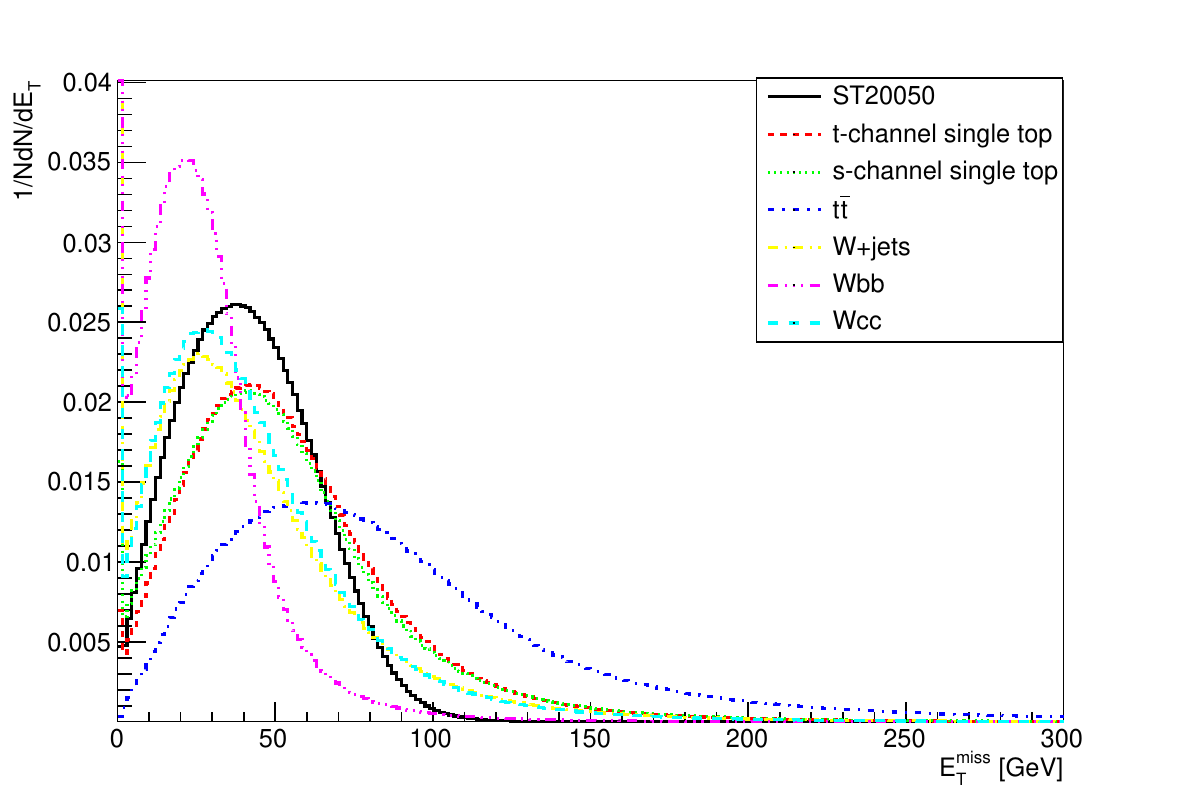}
   \caption{The missing transverse energy distribution is shown.}
   \label{met}
 \end{figure}

\begin{table}[h]
\begin{tabular}{|c|c|}
\hline
\hline
\hline
\hline
Isolation cut & $0.05<\Delta R(Leading trks, trks)<0.4$, $P_{T}^{trks}>1$ GeV \\
\hline
Ratio cut & $P_{T}^{max}/E_{T}^{jet}>0.3$, $P_{T}^{max}>4$\\
\hline
Top mass windSow & $150<m_{\tau \nu b}<190$ GeV \\
\hline
$\Delta \eta$ cut & $\Delta \eta (\tau-jet,b-jet) < 1$ \\
\hline
$\Delta \phi$ cut & $\Delta \phi (top,b-jet)>2.8$ \\
\hline
Missing $E_{T}$ & $E_{T}^{miss}>20$ GeV \\
\hline
\end{tabular}
\caption{The kinematical selection cuts applied on both signal and background events. \label{selectioncuts}}
\end{table}
 The Figure~\ref{bjets shows b-jet multiplicity for both signal and background events}. In each event, two such b-jets are required which fulfill the above criteria. In the signal events the b-jets distributions has slight dependence  on charged Higgs mass . In the heavy charged Higgs production and its decays to top quark the  production of b-jets is kinematically suppressed because of a smaller phase space availability .The selection of $\tau$-jets through its hadronic decays is the next phase in the reconstruction procedure of charged  Higgs .For  identification of $\tau$-jet , the standard algorithm  developed for LHC experiments  is used in this analysis. 
The important signature of $\tau$-jets through hadronic decay is the low charged particle multiplicity , where the highest leading tracks (high $P_{T}$) in the cone around $\tau-$lepton is supposed to be harder than the corresponding tracks in the light jets. Therefore to remove such soft $P_{T}$ tracks from the other light jets, the requirement of the leading tracks with $P_{T}$ $>$ 4 GeV is applied which makes the signal more cleaner and visible. The isolation requirement applied on $\tau$-jets is based on the fact that the $\tau$-jet accommodates only few charged tracks around the central axis of the jet cone. Therefore no charged tracks with $P_{T}$ above a certain threshold ($P_{T}$ $>$ 1 GeV) are expected to be in the annulus defined with cone size $0.05 < \Delta R < 0.4$. This cone annulus is defined in the $\tau$-jet cone around the leading tracks as shown in Figure~\ref{LeadingTPt} . Since τ$\tau$-jets preferably decay to either one or three charged pions, so one can expect one or three charged tracks in the signal cone defined as $\Delta R < 0.05$ around the leading track. The  Figure~\ref{nstracks} shows the distribution of the tracks present in the signal cone as well as tracks from background . Within the defined signal cone the number of signal tracks  are either one or three. So there should be only one jet satisfying all the above requirements of the $\tau$-identification algorithm. On the other hand , missing transverse energy is calculated as the negative sum over the particle momenta in horizontal and vertical directions.  Figure~\ref{met} shows the distribution  of missing transverse energy $E_{T}$   in signal and background events.For reconstruction of invariant mass  of W boson four-momentum ,tau-jet components and  $E_{miss}^{T}$ are used . At hadron colliders the z-component of the neutrino momentum is generally unknown  and thus is constructed by giving a right value for the W mass in the $\tau \nu$ combination, i.e.,$m_{\tau\nu} = $80 GeV.  If such a solution is not founded, $p^{\nu}_{z}$ is set to zero. A situation occurs in rare cases , in which  the mass of W candidate results different from the nominal value. Therefore It is not necessary to apply mass window on the W candidate invariant mass . Now the right b-jet from the top quark decay is found by calculating the top quark invariant mass using W boson four momentum through the $\tau\nu b$ combination and finding the b-jet which gives the closest top quark mass $m_{\tau \nu b}$ to the nominal value, i.e., $m_{top} = 173$ GeV. The $\Delta \eta(\tau$-jet, b-jet) cut which is plotted in Figure~\ref{deltaeta} helps to find such a true b-jet that is decaying from top quark. The $m_{\tau \nu b}$ distribution is shown in Figure~\ref{topmass} along with the surviving background processes. \\

\begin{figure}[tb]
 \centering
   \includegraphics[width=0.6\textwidth]{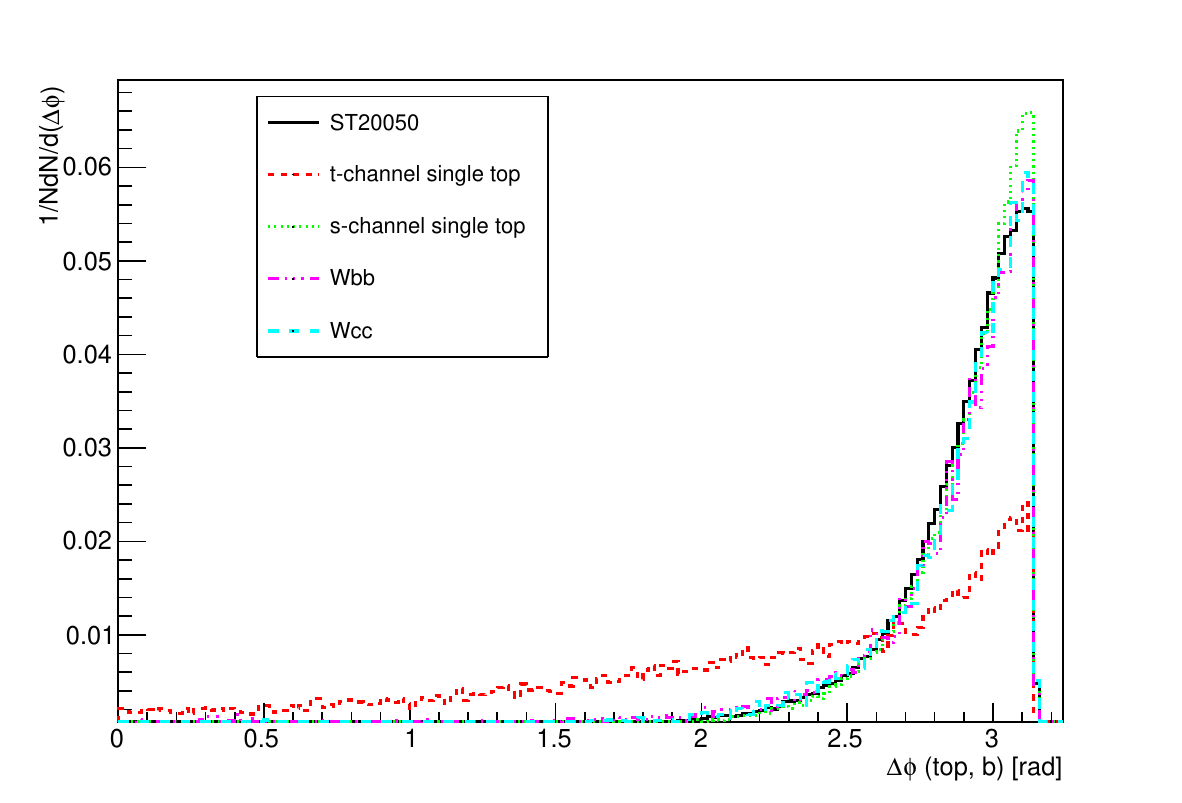}
   \caption{The difference in azimuthal angle between top and bottom quark is plotted. This cut helps to suppress those events which are not produced back-to-back.}
   \label{deltaphitb}
 \end{figure}

\begin{figure}[tb]
 \centering
   \includegraphics[width=0.6\textwidth]{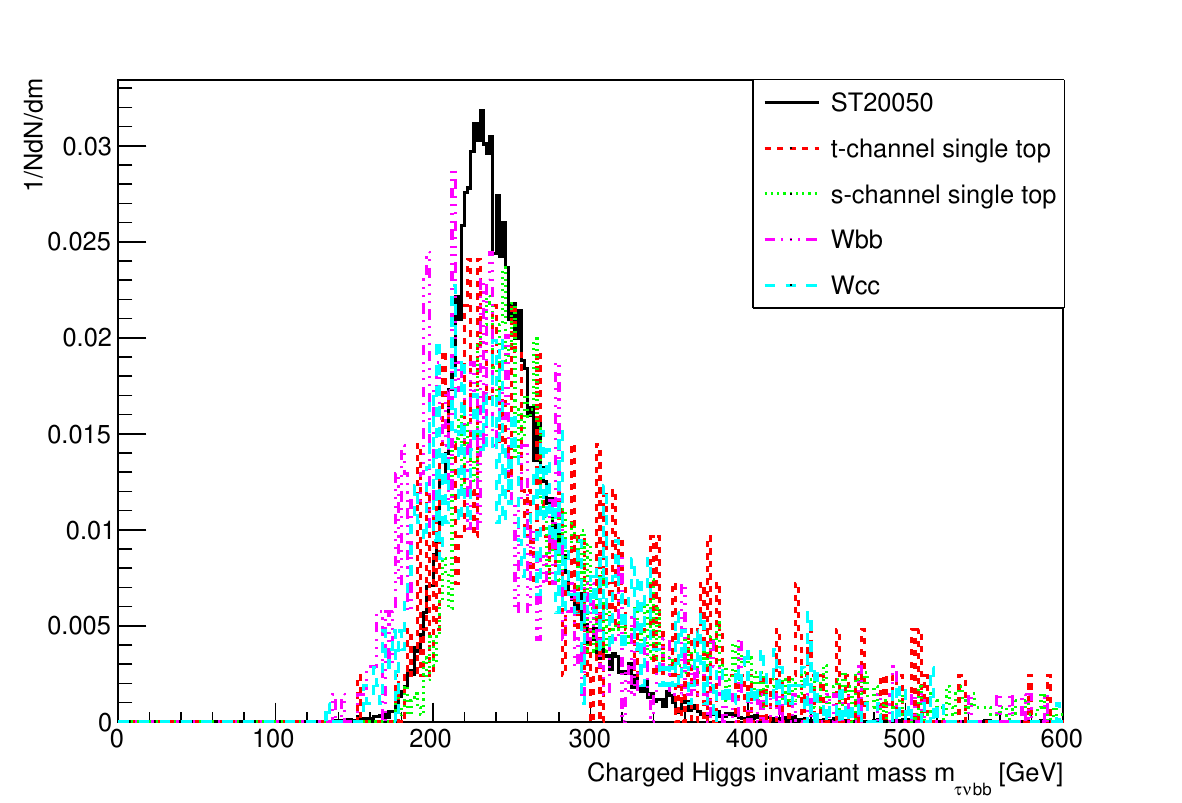}
   \caption{The reconstructed charged Higgs invariant mass distribution is obtained from $bb\tau\nu$ combination. Both signal and background events are normalized to unity to investigate the most probable process}
   \label{CHmass}
 \end{figure}

To reconstruct the charged Higgs boson candidate invariant mass a mass window of top quark invariant mass distribution is applied and another b-jet is combined with the top quark candidate e.g., $m_{\tau \nu bb}$. This is very interesting fact that signal events tend to produce back to back top and bottom quark in opposite direction and this feature appears in the azimuthal plane of the detector.To make the signal more dominant  on top of the background a cut is applied on the azimuthal angle between top and bottom as shown in Figure~\ref{deltaphitb}. Selection cut $E_{miss}^{T}$  is applied. Finally the charged Higgs candidate is reconstructed through $\tau\nu bb$ combination which is correlated as charged Higgs plotted in Figure~\ref{CHmass} and to suppress further background events a cut on missing $E_{T} > 20$ GeV is applied .\\

\begin{figure}[tb]
 \centering
   \includegraphics[width=0.6\textwidth]{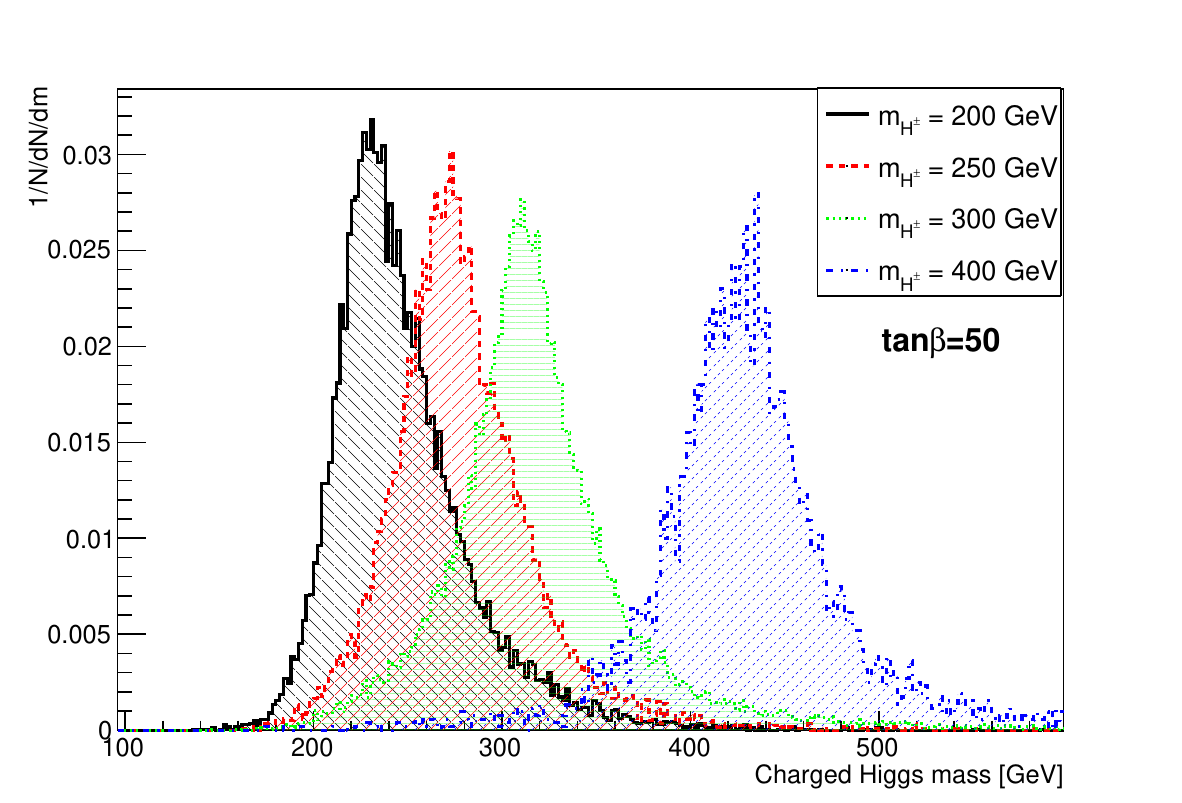}
   \caption{The reconstructed charged Higgs mass distributions at different input mass hypothesis.}
   \label{chmassall}
 \end{figure}
\begin{figure}[tb]
 \centering
   \includegraphics[width=0.6\textwidth]{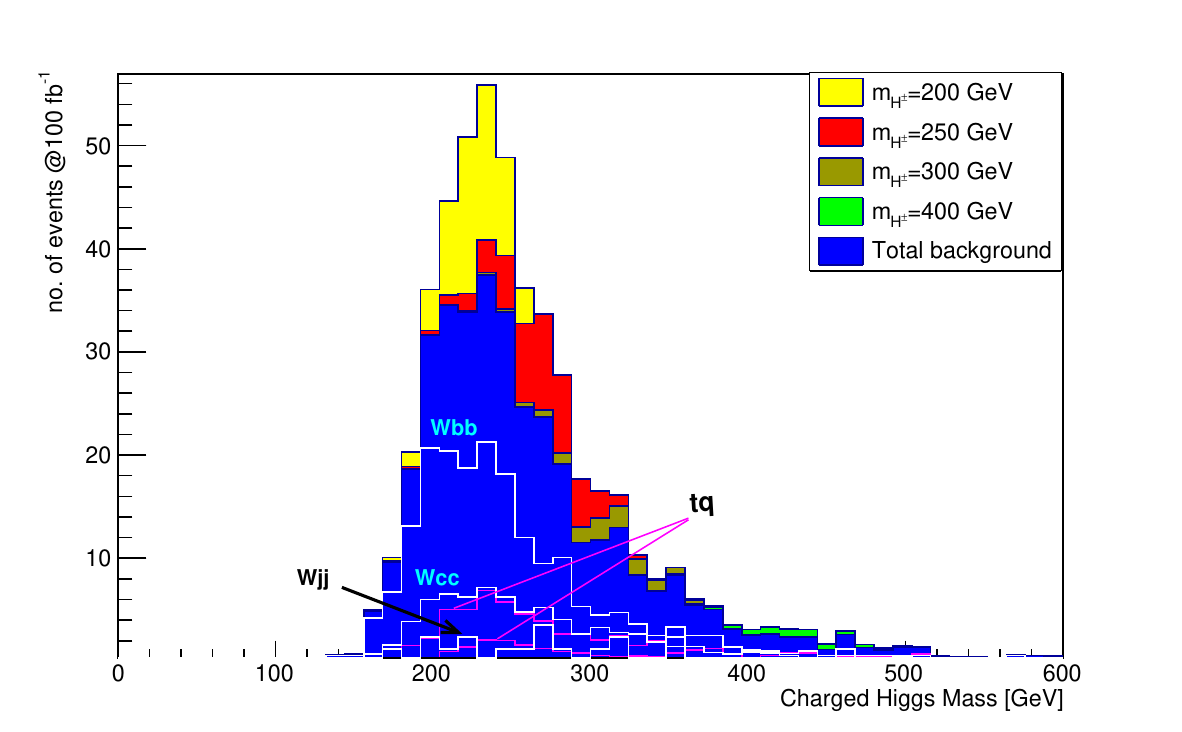}
 \caption{{Each sample is normalized to the real number of events obtained at 100 $fb^{-1}$}. The charged Higgs is on the top of the total background events at different charged Higgs mass hypotheses independently at tan$\beta = 50$. It shows its visibility for charged Higgs observability. Only dominant backgrounds are labeled.}
   \label{CHmassbins}
 \end{figure}

Now all signal and background events are allowed to pass through all above requirements separately and corresponding relative efficiencies are obtained with respect to previous cut as given in Table~\ref{signalcuts} for signal events and Table~\ref{backgroundcuts} for background events. The Table~\ref{signalcuts} corresponds to four charged Higgs mass hypotheses $m_{H^{\pm}}$ = 200, 250, 300, 400 GeV at tan$\beta=$ 50. As mass of charged Higgs  $m_{H^{\pm}}=180$ GeV  is close to top quark mass  and also it becomes hard  to observe decay of charged Higgs  into top and bottom quarks due to a very limited parameter space available so charged Higgs have not been considered . Usually this feature results in a soft kinematics of the final state particles. The $t\bar{t}$ events may also be treated as a source of charged Higgs bosons, when one of the top quark decays to charged Higgs, i.e., $t\bar{t}\rightarrow H^{\pm}W^{\pm}b\bar{b}\rightarrow \tau \nu jjb\bar{b}$. Such events are unlikely to be selected by the single top selection procedure and rejected badly. Consequently, the $t\bar{t}$ has no sizable contribution in addition of asking 3 jets in each event. The soft QCD multi jet sample is also entirely vanished. Taking $m_{H^{\pm}}$ = 200 GeV as an example the $\sigma \times BR(H^{\pm}\rightarrow tb$) is 5.63 $pb$ and a total efficiency of 0.0018 is obtained which leads to 1013 events at 100 $fb^{-1}$ being the largest statistics out of all signal hypotheses. In principle the invariant mass distribution of top and bottom quark should make the charged Higgs boson mass, however due to several effects e.g., jet energy resolution, false jet combination, mis-identification of jets and errors in their energy and flight directions create the shifts in the peaks of charged Higgs invariant mass around the input mass as can be observed in Figure~\ref{chmassall}. \\
Each obtained distribution is normalized with the real number of events at 100 $fb^{-1}$ including selection efficiencies as shown in Figure~\ref{CHmassbins}. Finally the number of signal and background events, efficiencies corresponding to different charged Higgs mass windows, signal to background ratio ($S/B$) and optimized signal significance ($S/\sqrt{B}$) are shown in Table~\ref{significance}. 

\begin{table}[h]
\begin{tabular}{|c|c|c|c|c|}
\hline
Selection Cut & Signal & Signal & Signal &Signal \\
& $M_{H^{\pm}}$ = 200 GeV & $M_{H^{\pm}}$ = 250 GeV & $M_{H^{\pm}}$ = 300 GeV & $M_{H^{\pm}}$ = 400 GeV \\
\hline
$\sigma$ x BR[$pb$] & 5.63 & 4.68 & 2.73 & 0.98  \\
\hline
3 jets & 32$\%$ &25$\%$ &21.7$\%$ &18.5$\%$ \\
\hline
2 b-jets &42$\%$ &56$\%$ &59.5$\%$ & 61.5$\%$ \\
\hline
Leading tracks &99.6$\%$ &99.8$\%$ &99.9$\%$ & 99.9$\%$ \\
\hline 
Isolation cut &$37\%$ &36.7$\%$ &36.6$\%$ &36.6$\%$ \\
\hline
Ratio cut &25$\%$ &25.7$\%$ &26$\%$ & 26.7$\%$ \\
\hline
no. of signal tracks & 97.8$\%$ &98$\%$ &98$\%$ & 97.8$\%$\\
\hline
1 tau jet &$99.5\%$ &99.8$\%$ &99.7$\%$ & 99.8$\%$\\
\hline
Top mass window &56.7$\%$ &44.4$\%$ &31$\%$ & 21$\%$\\
\hline
$\Delta \eta (\tau-jet,b-jet)$ &75.7$\%$ &75.6$\%$ &67$\%$ & 58.2$\%$\\
\hline
$\Delta \phi (top,b-jet)$ &$75\%$ &74$\%$ &80.6$\%$ & 89.8$\%$ \\
\hline
$E^{missing}_{T}$ &$46.8\%$ &39$\%$ &41.7$\%$ & 50$\%$\\
\hline
Total Efficiency &0.18$\%$ &0.13$\%$ &0.08$\%$ & 0.06\\
\hline
Expected events & 1013  & 608 & 218 & 59  \\
at 100 $fb^{-1}$ & & & &  \\
\hline
\end{tabular}
\caption{Selection efficiencies are shown for all signal events. \label{signalcuts}}
\end{table}

\begin{figure}[tb]
 \centering
   \includegraphics[width=0.6\textwidth]{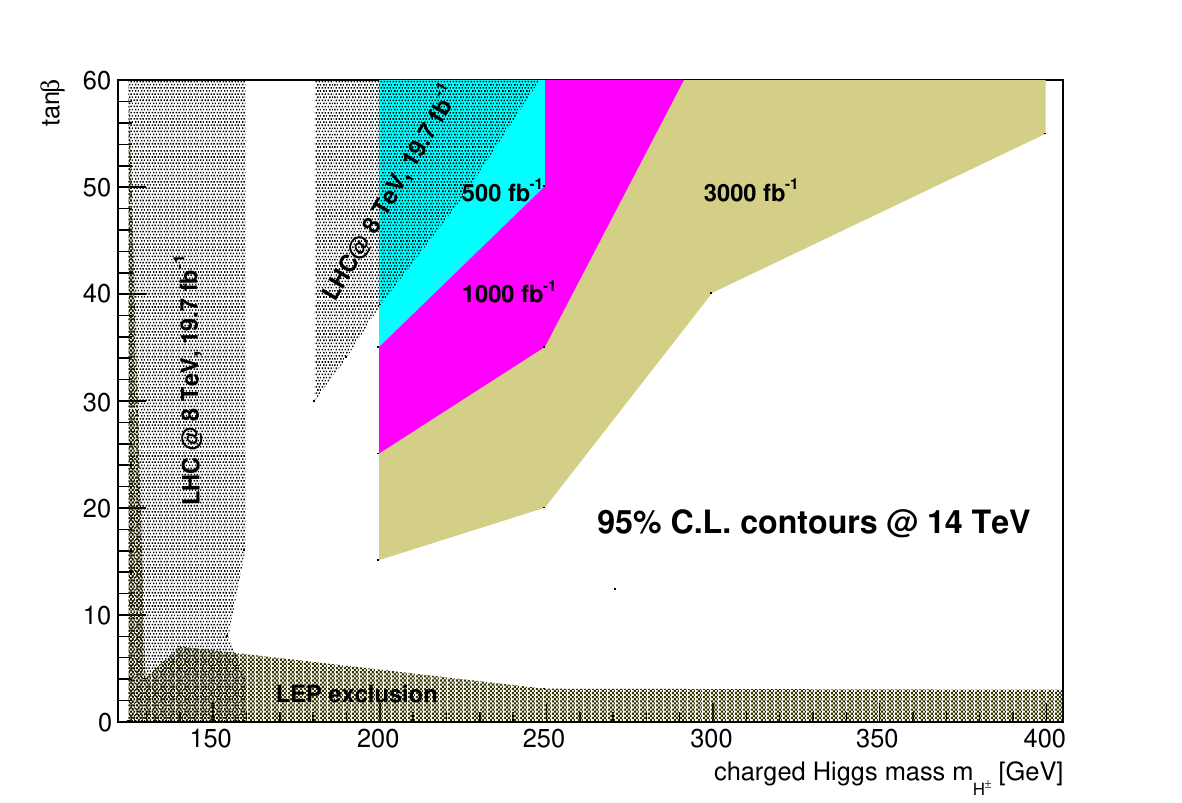}
 \caption{The exclusion curves with three colour bands shown, corresponds to 95$\%$ confidence level contour between $m_H^{\pm}$ and tan$\beta$.}
   \label{95cl}
 \end{figure}

\begin{figure}[tb]
 \centering
   \includegraphics[width=0.6\textwidth]{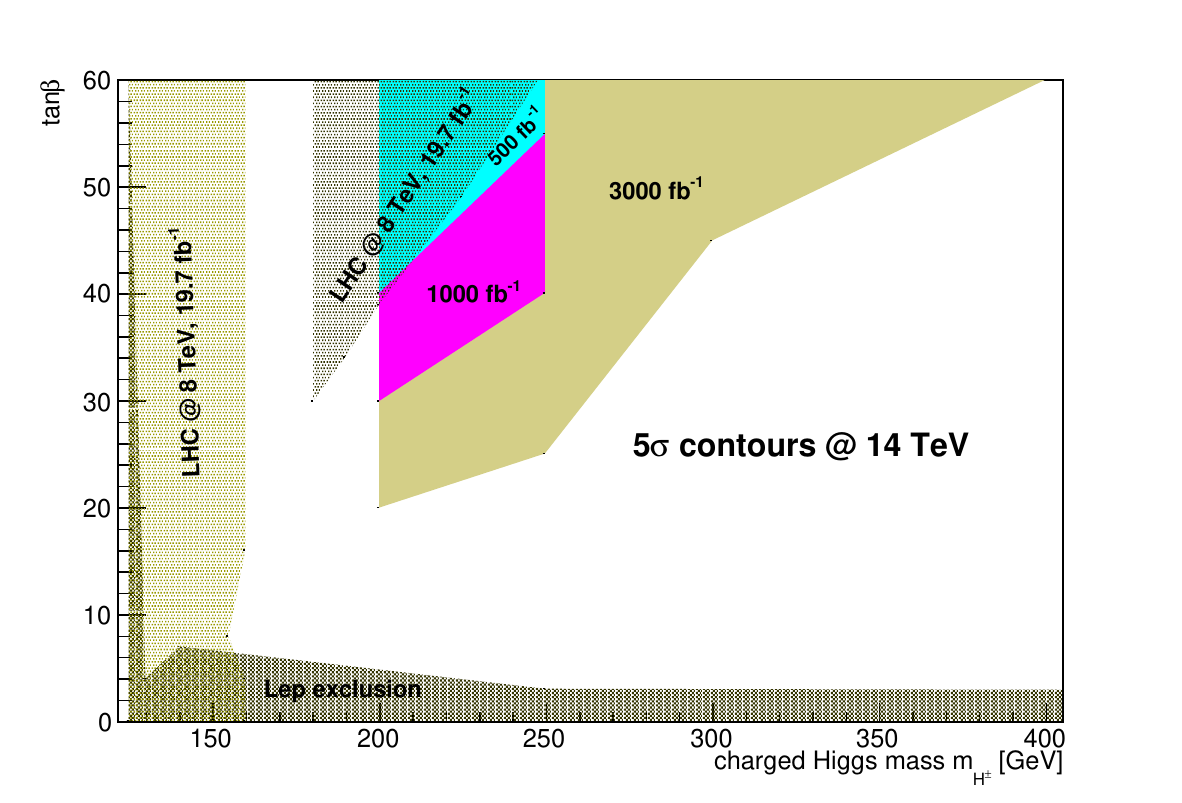}
  \caption{The phase space plot between $m_H^{\pm}$ (GeV) and tan$\beta$ is shown in order to demonstrate the 5$\sigma$ discovery contours at different integrated luminosities along with previous experimental exclusion curves.}
   \label{5sigma}
 \end{figure}

The charged Higgs mass window cut is applied in a specific region where a maximum signal significance is achieved which approaches to its best value of 32$\%$. This requirement significantly reduces background events e.g., the QCD jets are constrained due to jets-quark matching, even hough from  generation of million of events, few events could not survive. Now at the end we need to demonstrate the results validity in the MSSM parameter space by presenting all previous experimental constraints, 5$\sigma$ discovery contours and exclusion curves at 95$\%$ confidence level. These results are obtained by scanning the chosen charged Higgs mass points and tan$\beta$ values. On the basis of efficiencies obtained in Table~\ref{signalcuts} and Table~\ref{backgroundcuts}, the phase space exclusion limits are obtained at the 95$\%$ confidence level using a TLimit class implemented in ROOT \cite{root} package. These results are shown in Figures~\ref{95cl} and \ref{5sigma} incorporating previously excluded phase space areas by LEP, Tevatron and LHC data at 8 TeV. \\

\begin{table}[h]
\begin{tabular}{|c|c|c|c|c|c|c|}
\hline
Selection Cut & $t\bar{t}$ & SM single top & SM single top &Wjets& Wb$\bar{b}$ & Wc$\bar{c}$  \\
&  & s-channel & t-channel & & & \\
\hline
$\sigma$ x BR[$pb$] & 285.4 & 5.8 & 133 & 1.69$\times10^{4}$ & 395 & 49 \\
\hline
3 jets & 2.96$\%$ &25$\%$ &33$\%$ & 26$\%$&20 $\%$ &29 $\%$\\
\hline
2 b-jets &98$\%$ &41$\%$ &6.5$\%$ &0.31$\%$& 7$\%$ &13$\%$  \\
\hline
Leading tracks &51$\%$ &99.8$\%$ &99.6$\%$ &99.6$\%$ &99$\%$ &99.6$\%$ \\
\hline
Isolation cut &48$\%$ &28$\%$ &28$\%$ &26$\%$ &36$\%$ &36$\%$\\
\hline
Ratio cut &95$\%$ &18$\%$ &22$\%$ &18$\%$ &18$\%$ &18 $\%$\\
\hline
Number of signal tracks &52$\%$ &97$\%$ &97$\%$ &91$\%$ &97$\%$ &95 $\%$\\
\hline
1 tau jet &58$\%$ &99.7$\%$ &99.7$\%$ &99.7$\%$ &99$\%$ &99 $\%$\\
\hline
Top mass window &75$\%$ &42$\%$ &48$\%$ &32$\%$ &38$\%$ & 38$\%$\\
\hline
$\Delta \eta (\tau,b-jet)$ &59$\%$ &75$\%$ &69$\%$ &53$\%$ &60$\%$ &60 $\%$\\
\hline
$\Delta \phi (top,b-jet)$ &90$\%$ &74$\%$ &33$\%$ &49$\%$ &68$\%$ &68 $\%$\\
\hline
$E^{missing}_{T}$ &49$\%$ &46$\%$ &43$\%$ &60$\%$ &46$\%$ &46 $\%$\\
\hline
Total Efficiency &0.05$\%$ &0.05$\%$ &0.006$\%$ &0.00018 $\%$ &0.007 $\%$ &0.017$\%$ \\
\hline
Expected events &14280  & 290 &798 &30420  &2765  &833  \\
at 100 $fb^{-1}$ & & & & & & \\
\hline
\end{tabular}
\caption{Selection efficiencies are shown for all background events. \label{backgroundcuts}}
\end{table}

\begin{table}[h]
\begin{tabular}{|c|c|c|c|c|c|c|}
\hline
Sample &\multicolumn{2}{c}{Mass window Gev} & Total & no. of & S/B & Optimized  \\
& Lower limit & Upper limit & Efficiency & Events & &S/$\sqrt{B}$ \\
\hline
Signal, $m_{H^{\pm}}$ = 200 GeV & 114 & 150 & 0.0014 & 411 & 0.32 & 11\\
 Total Background & 114 & 150 &  & 1266 & & \\
\hline
Signal, $m_{H^{\pm}}$ = 250 GeV & 126 & 198 & 0.001 & 250 & 0.197 & 7\\
Total Background & 126 & 198 &  & 1276 & & \\
\hline
Signal, $m_{H^{\pm}}$ = 300 GeV& 162 & 294 & 0.0005 & 72 & 0.132 & 3  \\
 Total Background & 162 & 294 &  & 540 &  & \\
\hline
Signal, $m_{H^{\pm}}$ = 400 GeV& 216 & 294 & NS  & 15 & 0.09 & 1\\
Total Background & 216 & 294 &  &  165 &  &  \\
\hline
\end{tabular}
\caption{Signal to background ratio and signal significance values obtained for four different samplesat 100 $fb^{-1}$ , where "NS" represents negligibly small.\label{significance}}
\end{table}

\section{Conclusion}
The charged Higgs observability potential is investigated in the single top s-channel process and proved a source of charged Higgs boson in $\tau$ fully hadronic decay at $\sqrt{s}=$14 TeV at LHC. The Optimized kinematical selection cuts result in enhancing the signal to background ratio as well as signal significance specifically at phase space ($m_H^{\pm}$ = 200 GeV, tan$\beta$ = 50). On the basis of presented results of exclusion curves at 95$\%$ confidence level and 5$\sigma$ contours, it may be concluded that the charged Higgs signal can be well observed or excluded in a wide range of phase space particular (tan$\beta > 35$ at 500 $fb^{-1}$, tan$\beta > 25$ at 1000 $fb^{-1}$, tan$\beta > 15$ at 3000 $fb^{-1}$). An important source of ambiguities and uncertainties are the systematic errors and theoretical uncertainties which have not been taken into account in this study. Undoubtedly, in the realistic approach at hadron colliders experiments, one needs to incorporate all the sources of uncertainties e.g., pile-up, electronic noise, trigger inefficiencies, acceptance and vertex reconstruction related errors etc. Such analysis where the final state is jets($b's$ and $\tau's$) , the expected dominant source of uncertainty is the jet energy resolution which may be less than 1$-$2 $\%$ in the central region of the detector for jets having transverse momentum in the range of 55-500 GeV \cite{jetscale}.

\end{document}